\documentclass[
twocolumn,
prl,showpacs,preprintnumbers,amsmath,amssymb]{revtex4}

\usepackage{graphicx}
\usepackage{amssymb}
\usepackage{amsmath}
\usepackage{dsfont}
\usepackage{bm}
\usepackage{mathrsfs}

\usepackage{graphicx}

\begin{document}

\title{Conformal cloak for waves}
\author{Huanyang Chen$^1$, Ulf Leonhardt$^2$, and Tom\'{a}\v{s} Tyc$^{3}$}
\affiliation{
$^1$ School of Physical Science and Technology, Soochow University, Suzhou, Jiangsu 215006, China\\
$^2$School of Physics and Astronomy, University of St Andrews,
North Haugh, St Andrews KY16 9SS, UK\\
$^3$Faculty of Science, Kotlarska 2, and Faculty of Informatics, Botanicka 68a,
Masaryk University, 61137 Brno, Czech Republic}
\date{\today}
\begin{abstract}
Conformal invisibility devices are only supposed to work within the validity range of geometrical optics. Here we show by numerical simulations and analytical arguments that for certain quantized frequencies they are nearly perfect even in a regime that clearly violates geometrical optics. The quantization condition follows from the analogy between the Helmholtz equation and the stationary Schr\"odinger equation.
\end{abstract}
\maketitle

Invisibility came into sight as the first nontrivial application of transformation optics \cite{Greenleaf,LeoConform,PSS,CCS,Book}. The progress made has been impressive, but complete cloaking devices have never been demonstrated in practice yet. All electromagnetic cloaking experiments reported in the literature were for reduced cases. For example, the first cloaking device \cite{Schurig} worked for microwaves of one frequency and polarization, and this only in the approximation of geometrical optics. Cloaking in tapered waveguides is also approximative \cite{Waveguides}. Carpet cloaking \cite{LiPendry,Carpet} is a drastic form of reduced cloaking where an object is not made to disappear completely, but to appear as being flat. There are two reasons for the need of resorting to reduced cloaking devices, a practical and a fundamental one. The materials required for perfect cloaking \cite{Greenleaf,PSS} are extremely difficult to fabricate, because they need to implement impedance-matched anisotropic media \cite{Book}, and therefore perfect cloaking is impractical. The fundamental problem is that perfect cloaking also implies propagation with a superluminal phase velocity that reaches infinity \cite{GREE}, which is possible in principle, but only for discrete frequencies. Perfect cloaking in a broad band of the spectrum is therefore physically impossible. Note that these qualifications do not apply to acoustic cloaking \cite{ChenChan,ZhangXiaFang} where the required materials are much easier to manufacture and the constraints from relativistic causality are not relevant.

One of the earliest ideas for invisibility devices, Optical Conformal Mapping \cite{LeoConform,LeoNotes}, has the advantage of requiring rather ordinary optical materials --- they are optically isotropic and nonmagnetic (but still need a large index range). It can be easily applied in acoustics, because it does not rely on materials with artificially designed anisotropic mass \cite{ChenChan,ZhangXiaFang}. This cloaking method was originally derived \cite{LeoConform} for the approximation of geometrical optics (or acoustics). In isotropic media, perfect invisibility is mathematically impossible, not only for physical reasons, because the inverse scattering problem is uniquely solvable \cite{Nachman} and so waves cannot completely hide from the fact that they propagate in media. Here we show, however, that conformal cloaking can be nearly perfect for discrete frequencies.

Cloaking by Optical Conformal Mapping consists of two parts: (1) the implementation of a conformal coordinate transformation and (2) an index profile with certain properties that cannot be reduced to flat space by a coordinate transformation, a non-Euclidian part \cite{LeoTyc,Tyc}. A conformal coordinate transformation is valid for complete waves, but it cannot make a cloaking device on its own \cite{LeoConform}; the non-Euclidean part completes the device, but it is only supposed to work within the validity range of geometrical optics. Full-wave simulations of Optical Conformal Mapping are rare; in a recent review \cite{Urzhumov} the effect of conformal coordinate transformations was studied, but without the non-Euclidean part. Not surprisingly, cloaking is not possible here, but it is wrong to conclude from an incomplete numerical experiment that cloaking by Optical Conformal Mapping would not work. The simulations we show here prove to the contrary, even in a regime far away from geometrical optics.

Let us begin with a brief recapitulation of Optical Conformal Mapping \cite{LeoConform,LeoNotes}. Consider a planar medium with the two-dimensional graded index profile $n(x,y)$. The medium is purely electrical such that $\varepsilon=n^2$. In this case, electromagnetic waves polarized such that the electric field points in vertical direction (orthogonal to the plane) obey the Helmholtz equation \cite{LiPendry}
\begin{equation}
0 = \left(\partial_x^2+\partial_y^2 + n^2k^2\right)\psi
= \left(4\partial_z^*\partial_z + n^2k^2\right)\psi 
\label{helmholtz}
\end{equation}
where $\psi$ denotes the electric-field component, $z=x+\mathrm{i}y$ and $k$ is the wavenumber. Electromagnetic waves polarized such that the magnetic field points in vertical direction also obey the Helmholtz equation (\ref{helmholtz}), but only approximately in the regime of geometrical optics \cite{Book}. In this case $\psi$ denotes the magnetic-field component. A linear combination of the two polarizations constitutes an arbitrary electromagnetic wave in the planar medium. Under a conformal mapping $w=w(z)$ the Helmholtz equation in $w$ space appears as
\begin{equation} 
\left(4\partial_w^*\partial_w + n'^2k^2\right)\psi = 0 \,,\quad
n = n' \left|\frac{\mathrm{d}w}{\mathrm{d}z}\right| \,.
\label{whelmholtz}
\end{equation}
In the language of transformation optics \cite{CCS,Book} $w$-space is called virtual space and  $z$-space is the physical space. Consider the simplest mapping suitable for cloaking, the Zhukowski transform \cite{LeoConform,Book}
\begin{equation} 
w = z+\frac{a^2}{z} \quad\mbox{or, equivalently,}\quad z = \frac{w\pm\sqrt{w^2-4a^2}}{2}
\label{zhu}
\end{equation}
that maps a virtual $w$-space with two Riemann sheets on two regions of physical $z$-space, one is the exterior and the other the interior of a circle of radius $a$ (Fig.~1). If the virtual space were empty light rays would travel along straight lines in $w$-space that may cross the branch cut from the exterior to the interior $w$ sheet and then get absorbed at the infinity of the latter that corresponds to the singularity of $w(z)$, which, for the Zhukowski transform (\ref{zhu}), lies at $z=0$ in physical space. This feature has caused the distortion of wave propagation \cite{Urzhumov} mentioned above. However, it is possible to shepherd the lost rays back to the exterior sheet by filling the interior sheet with a suitable index profile that cannot be reduced to empty space by a coordinate transformation \cite{LeoConform,LeoNotes}. Many of such profiles are possible \cite{Book}, including ones with negative refraction \cite{Ochai} (that do not cause phase delays). What these index profiles have in common is their ability to let light propagate in closed trajectories. Consider the simplest case, the Hooke and the Kepler profile \cite{LeoConform,LeoNotes}:
\begin{eqnarray} 
n'^2 &=& 1 - \frac{|w-w_1|^2}{r_0^2} \quad \mbox{(Hooke)}
\label{hooke} \\
n'^2 &=& \frac{r_0}{|w-w_1|} - 1 \quad \mbox{(Kepler)}
\label{kepler}
\end{eqnarray}
where $w_1$ is the branch point $2a$ of the Zhukowski map (\ref{zhu}) and $r_0$ is chosen to be $4a$. The index $n$ in physical space ranges from $0$ to $14.5$ for the cloaking device with Hooke profile and from $0$ to $13.3$ for the Kepler case. In both cases, the ray trajectories are closed curves in virtual space (in particular ellipses) such that after a loop in the ``underworld'' --- on the interior sheet --- the light returns to the exterior sheet and propagates along straight lines in $w$-space in the same direction it came from. In the ``underworld'' the light is confined to a circle around $w_1$ with radius $r_0$, because beyond this circle $n'$ is purely imaginary and so light cannot propagate there (light waves decay exponentially). In physical space, the exterior of the virtual circle is the interior of its map $z(w)$, i.e.\ of the curve $r=a\left(2+\cos\phi-\sqrt{(2+\cos\phi)^2-1}\,\right)$ in polar coordinates. This region --- and anything inside it --- is invisible, while light bends around it such that rays are asymptotically straight lines in physical space: the conformal transformation (\ref{zhu}) combined with one of the profiles (\ref{hooke}) and (\ref{kepler}) makes a cloaking device. Note that light rays are refracted at the interface between the interior and the exterior region of the Zhukowski map (\ref{zhu}) where the profiles (\ref{hooke}) and (\ref{kepler}) reside. Rays are refracted back to their original direction after one loop on the interior sheet, but waves are partially reflected at sudden index steps. When the index varies much more rapidly over the scale of the wavelength, as it is the case for index steps, the conditions of geometrical optics \cite{Book} are violated, causing the conversion of one wave front into two, {\it i.e.} partial reflection. Furthermore, the detour in the ``underworld''  causes a uniform time delay \cite{LeoNotes} that appears as a phase shift in the part of the wave that entered there, and thus creates phase dislocations that may cause diffraction overshadowing the cloaking of small devices.
\begin{figure}[t]
\begin{center}
\includegraphics[width=18.0pc]{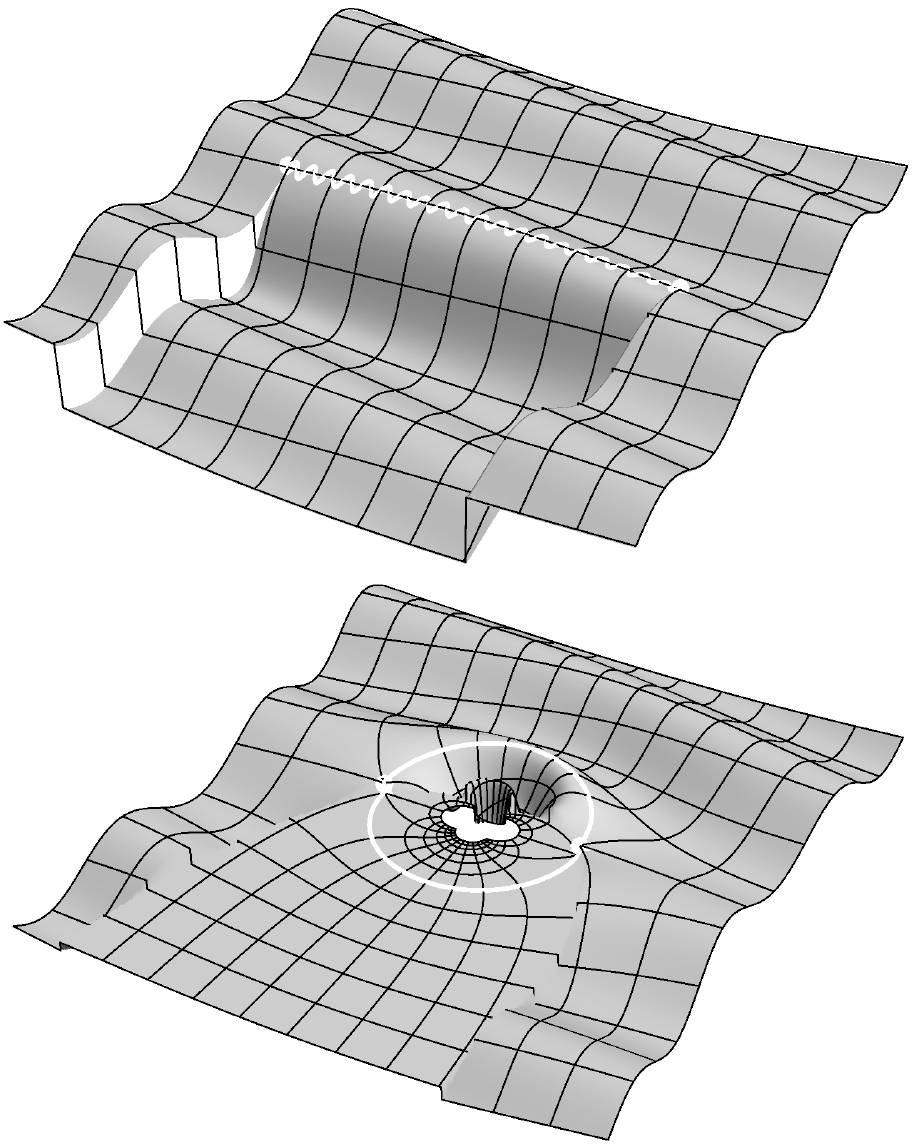}
\caption{
\small{
In transformation optics, electromagnetic waves are transformed from an empty virtual space to physical space by an appropriate medium. In Optical Conformal Mapping this is achieved by a conformal transformation that requires an optically isotropic material. The virtual space consists of Riemann sheets (above) that are mapped to the physical plane (below). However, light may cross the branch cut between two sheets and get absorbed at a singularity of the transformation, casting a shadow with zero amplitude, as the figure illustrates for the Zhukowski map (\ref{zhu}). Cloaking with optically isotropic materials (Fig.~2) is possible when the lower sheet contains a medium where light propagates in closed loops (Fig.~3).}
}
\end{center}
\vspace*{-7mm}
\end{figure}

Here we have tested the performance of conformal cloaking devices for waves using simulations made with standard commercial software. In the invisible region of the device we used an imaginary index profile. One can also shield this region with perfect mirrors; our simulations show little principal difference between the two (only quantitative differences). We also introduced a cut-off radius $r_c$ in physical space that makes our cloaking device finitely extended. Conformal transformations such as the Zhukowski map (\ref{zhu}) act across the entire plane, in contrast to quasiconformal transformations \cite{LiPendry}, so the medium implementing the transformation is, in principle, infinitely extended. Yet beyond the cut-off radius $r_c=5a$ we put $n=1$ and observed that this cut-off has a negligible effect on the propagation of waves. More importantly, our simulations show that the sharp index steps at the interface of the inner cloaking region do neither cause reflections nor phase delays for certain discrete wavenumbers of light (Fig.~2). 

\begin{figure}[h]
\begin{center}
\includegraphics[width=20.0pc]{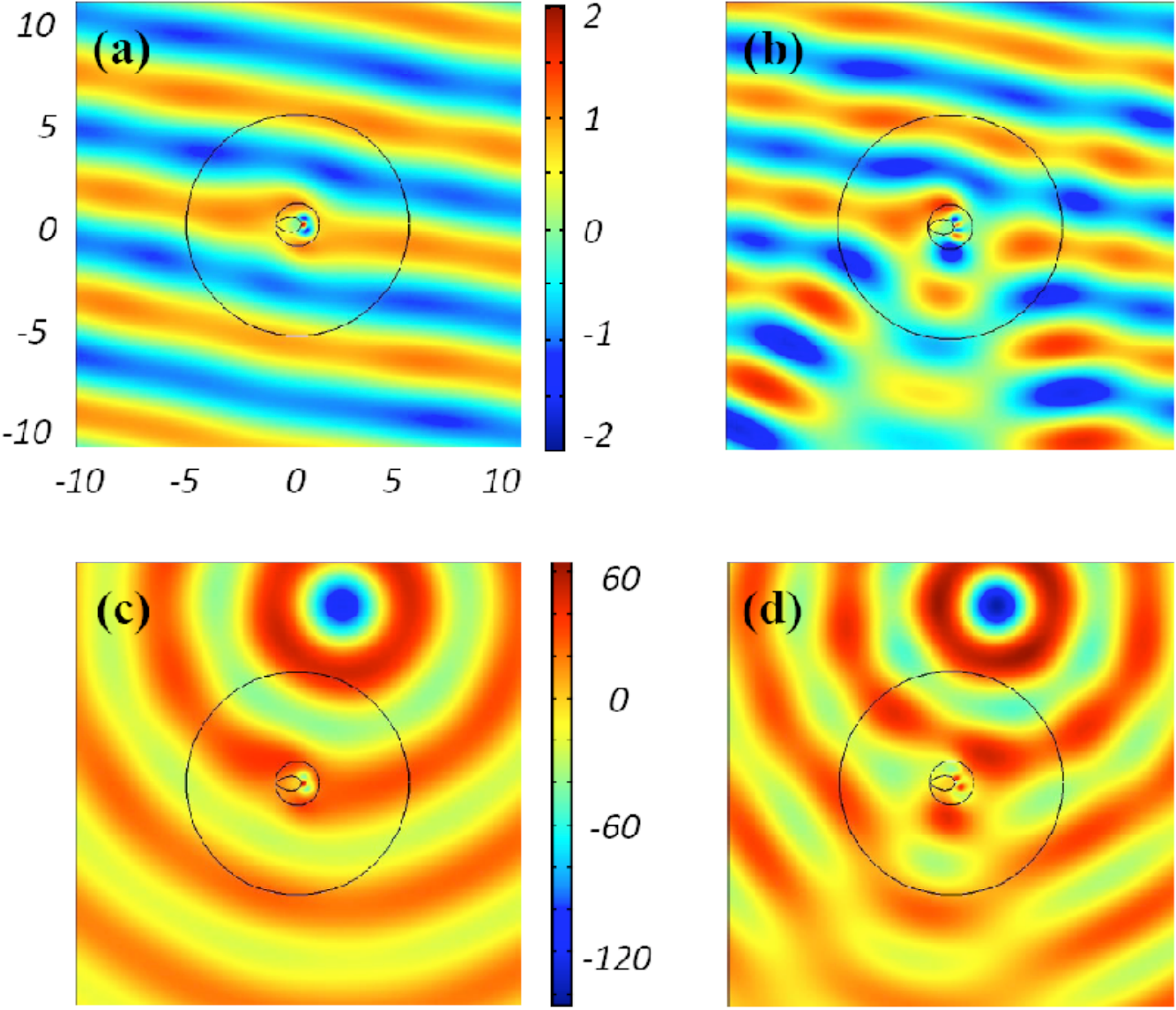}
\caption{
\small{
Color online. Simulations of wave propagation for a cloaking device based on the Zhukowski map (\ref{zhu}) and the Kepler profile (\ref{kepler}); our results for the Hooke profile (\ref{hooke}) are very similar. In each panel the outer circle describes the boundary of the device (with cut-off radius $r_c$), the inner circle contains the core of the cloaking device that carries the transformed profile (\ref{kepler}) and the pupil indicates the cloaked region where the wave decays exponentially. The top pictures (a,b) show the cloaking of incident plane waves and the bottom pictures (b,d) cylindrical waves. Our simulations illustrate two extreme cases, $kr_0=5$ (a,c) with nearly perfect invisibility and $kr_0=6$ (b,d) with a pronounced phase dislocation and a resulting diffraction pattern. Note that the core of the device is small in comparison with the wavelength such that geometrical optics is no longer a good approximation, yet cloaking can be nearly perfect.}
}
\end{center}
\end{figure}

The wavenumbers of nearly-perfect invisibility turn out to be related to the eigen-frequencies of light in the index profiles (\ref{hooke}) and (\ref{kepler}). We can easily deduce them from the analogy between the Helmholtz equation (\ref{whelmholtz}) in virtual space and the stationary Schr\"odinger equation 
\begin{equation} 
\left(\nabla'^2 +\frac{2m}{\hbar^2}(E-U)\right)\psi = 0 \,,\quad
E-U = \frac{\hbar^2n'^2k^2}{2m}
\label{schroedinger}
\end{equation}
where $U$ denotes the potential, $E$ the energy and $m$ the mass. The Hooke profile (\ref{hooke}) corresponds to the 2D harmonic-oscillator potential
\begin{equation} 
U = \frac{\hbar^2k^2}{mr_0^2}\,\left|w-w_1\right|^2 = \frac{m\omega_0^2}{2}\,\left|w-w_1\right|^2
\label{u}
\end{equation}
with $\omega_0$ being the oscillation frequency. An eigenstate of the 2D oscillator has the quantized energy
\begin{equation} 
E = \frac{\hbar^2k^2}{2m} = \hbar\omega_0(l+1)
\label{e}
\end{equation}
for non-negative integer $l$. From relations (\ref{u}) and (\ref{e}) follows 
\begin{equation} 
kr_0 = 2(l+1) \quad \mbox{(Hooke)}
\label{hookek}
\end{equation}
that defines the wavenumber $k$ of an eigenmode. Along similar lines we deduce the eigenmodes in the Kepler profile (\ref{kepler}) from the 2D hydrogen spectrum \cite{ZZ}, and obtain in this case
\begin{equation} 
kr_0 = 2l+1 \quad \mbox{(Kepler).}
\label{keplerk}
\end{equation}
It is clear from semiclassical quantum mechanics that for these eigenmodes the phase difference along the closed loop on the interior sheet in virtual space is an integer multiple of $2\pi$. Therefore the device does not cause a phase dislocation; the emerging wavefront on the exterior sheet is intact. What seems more surprising is the fact that the incident wave is not reflected at the boundary between exterior and interior sheet (within the accuracy of our simulations) even when the index jumps by an infinite amount, as is the case for the Kepler profile (\ref{kepler}) in virtual space. Finally, we observed (Fig.~2) that the cloaking device performs nearly perfectly for the wavenumbers (\ref{hookek}) and (\ref{keplerk}), showing no signs of reflection and diffraction beyond free wave propagation. Cloaking is almost perfect even when the device is smaller than the wavelength of light, a regime that lies outside the validity range of geometrical optics \cite{Book}.

\begin{figure}[t]
\begin{center}
\includegraphics[width=20.0pc]{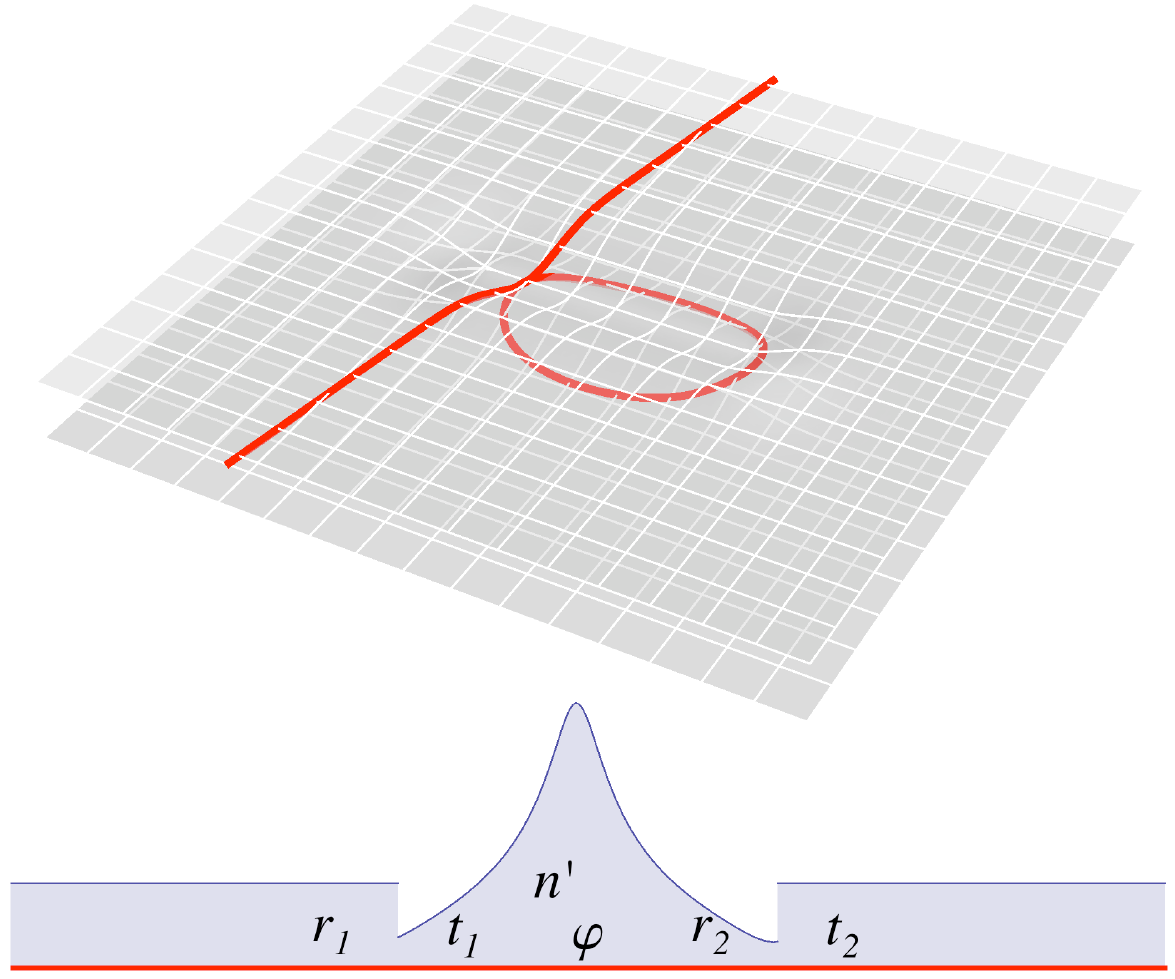}
\caption{
\small{
Color online. One-dimensional model. The top picture shows a typical closed loop of a light ray on the Riemann surface of the Zhukowski map (\ref{zhu}) with the Kepler potential (\ref{kepler}) on the lower sheet. The bottom picture visualizes the refractive index profile $n'$ experienced by the ray on its trajectory in virtual space. At the sharp interfaces between the medium a wave is reflected and transmitted with coefficients $r_i$ and $t_i$. However, when the phase delay $\varphi$ in the dielectric structure is an integer multiple of $2\pi$ no net reflection occurs due to interferences at multiple reflections. In this case the structure is perfectly invisible.}
}
\end{center}
\vspace*{-7mm}
\end{figure}

The perfect transmission and hence perfect invisibility at an eigenfrequency can be made plausible by a simple one-dimensional model. Imagine, instead of the lower Riemann sheet (Fig.~1) where the light enters and leaves through the branch cut, a one-dimensional index profile (Fig.~3) that light enters from one side and leaves at the other. This simple model contains the essence of the more complicated behavior of light propagation on the interior Riemann sheet where the index jumps, but after one loop returns to the original value; in the model we have opened this loop such that the two sides of the branch cut are represented by two interfaces. We denote the reflectivities and transmittivities of each interface by $r_1, r_2$ and $t_1, t_2$, and obtain from the Fresnel coefficients \cite{Book} the relations
\begin{equation} 
r_1 = -r_2 \,,\quad t_1t_2 = 1-r_1^2 = 1-r_2^2 \,.
\label{fresnel}
\end{equation}
Consider a wave $\psi$ propagating in $x$ direction in a uniform background medium with index $n_0$ towards the non-uniform profile representing the interior sheet. Part of the wave is reflected, the transmitted part propagates to the second interface with phase $\varphi$ and is multiply reflected inside the index profile, such that in total
\begin{eqnarray} 
\psi &=& \mathrm{e}^{\mathrm{i}n_0kx} +  \mathrm{e}^{-\mathrm{i}n_0kx}\left(r_1+t_1\,\mathrm{e}^{\mathrm{i}\varphi}\,r_2\,\mathrm{e}^{\mathrm{i}\varphi}\rho\, t_2\right) \,, \label{wave}\\
\rho &=& \sum_{l=0}^\infty \left(r_2\mathrm{e}^{\mathrm{i}\varphi}\right)^2 = \frac{1}{1-\left(r_2\mathrm{e}^{\mathrm{i}\varphi}\right)^2} \,.
\end{eqnarray}
For an eigenmode the phase delay $\varphi$ is an integer multiple of $2\pi$. In this case the Fresnel relations (\ref{fresnel}) imply that $\psi=\exp(\mathrm{i}n_0kx)$; the interference of light by multiple reflections causes no reflection at all --- the wave is perfectly transmitted and so the index profile is invisible.  It is remarkable that this behavior carries over to the wave propagation on virtual Riemann sheets that, for certain frequencies, are not only invisible themselves but make everything inside invisible.

We thus showed by numerical simulations and simple analytical arguments that cloaking devices based on Optical Conformal Mapping \cite{LeoConform} work perfectly for discrete frequencies --- like the cloaking devices made by the implementation of non-conformal coordinate transformations \cite{PSS} --- and this even in a regime far beyond geometrical optics. 

{\it Acknowledgements.---} H.C.\ is supported by grant 11004147 of the National Natural Science Foundation of China and grant BK2010211 of the Natural Science Foundation of Jiangsu Province. T.T.\ acknowledges the grants MSM 0021622409 and MSM 0021622419. U.L.\ is supported by the Royal Society.
 


\end{document}